\documentclass[apj,twocolumn]{openjournal}
\usepackage[dvipsnames]{xcolor} 
\usepackage[breaklinks,colorlinks,citecolor=blue,urlcolor=blue]{hyperref}
\usepackage{newtxtext,newtxmath}
\usepackage[T1]{fontenc}
\usepackage{ae,aecompl}
\usepackage{graphicx}
\usepackage{dcolumn}
\usepackage{bm}
\usepackage{color}
\usepackage{natbib}
\usepackage{multirow}\usepackage{orcidlink}

\usepackage{xspace}
\usepackage{nicefrac}

\newcommand{\B}[2]{\ensuremath{[\text{#1}/\text{#2}]}\xspace}
\newcommand{\Msun}{\ensuremath{\mathrm{M}_\odot}}

\newcommand{\ch}{\ensuremath{\mathrm{M_{Ch}}}\xspace}

\newcommand{\cm}{\ensuremath{\mathrm{cm}}}

\newcommand{\erg}{\ensuremath{\mathrm{erg}}}
\newcommand{\g}{\ensuremath{\mathrm{g}}}

\defcitealias{Thibodeaux2024}{T24}
\defcitealias{xing2023Natur}{X23}
\defcitealias{skuladottir2024ApJ}{S24}

\begin{document}

\title{Origin of LAMOST J1010+2358 Revisited}

\author{S. K. Jeena\orcidlink{0000-0002-6517-7419}}
\author{Projjwal Banerjee\orcidlink{0000-0002-6389-2697}}
\affiliation{Department of Physics\\
Indian Institute of Technology Palakkad, Kerala, India}
\email{Corresponding author: jeenaunni44@gmail.com}
\begin{abstract}
Signature from Pop III massive stars of $140$--$260\,{\rm M_\odot}$ that end their lives as pair-instability supernovae (PISNe) are expected to be seen in very metal-poor (VMP) stars of ${\rm [Fe/H]}\leq -2$. Although thousands of VMP stars have been discovered, the identification of a VMP star with a PISN signature has been elusive. Recently, the VMP star LAMOST J1010+2358 was claimed to be the first star with a clear PISN signature. A subsequent study showed that ejecta from low-mass core-collapse supernovae (CCSNe) can also fit the abundance pattern equally well and additional elements such as C and Al are required to differentiate the two sources. Follow-up observations of LAMOST J1010+2358 by two independent groups were able to detect both C and Al. Additionally, key odd elements such as Na and Sc were also detected whose abundances were found to be higher than the upper limits found in the original detection. We perform a detailed analysis of the newly observed abundance patterns by exploring various possible formation channels for VMP stars. We find that purely low-mass CCSN ejecta as well as the combination of CCSN and Type 1a SN ejecta can provide an excellent fit to the newly observed abundance pattern. Our results confirm earlier analysis that the newly observed abundance pattern is peculiar but has no signatures of PISN.

\end{abstract}


\section{Introduction}
The nature of the very first stars (Pop III) is one of the key questions in Astrophysics. Simulations of star formation in the early universe suggest that the initial mass function (IMF) for Pop III stars was top-heavy and dominated by massive stars of $\gtrsim 10\,\Msun$~\citep{abel1998first, Abel2000ApJ, Hirano2015MNRAS}. A consequence of this is that a substantial fraction of stars could have been in the mass range of $\sim 140\hbox{--}260\,\Msun$ that end their life as pair-instability supernovae (PISNe) \citep{ober1983,hegerwoosley2002}. Because the abundance pattern resulting from PISN ejecta is distinct from core-collapse supernovae (CCSNe) from regular massive stars of $\lesssim 100\,\Msun$, clear signatures from PISN are expected to be seen in some of the low-mass very metal-poor (VMP) stars with $\B{Fe}{H}\leq-2$ that formed with the first $\sim 1\,{\rm Gyr}$ from the Big Bang. However, even after the discovery of numerous VMP stars, no such star with a clear PISN signature has been found.

In this regard, \citet{xing2023Natur} (hereafter \citetalias{xing2023Natur}) recently discovered a chemically peculiar VMP star LAMOST J1010+2358 (hereafter J1010+2358) which was claimed to have a clear chemical signature of a PISN. 
However, an analysis by \citet{jeena_CCSN2024} found that although PISN does provide a very good fit to the observed pattern in J1010+2358, an equally good fit is provided by low-mass CCSNe of $12\hbox{--}14\,\Msun$ that undergo limited fallback (see also \citet{Koutsouridou2024ApJ}). It was pointed out by \citet{jeena_CCSN2024} that although both PISN and CCSN can match the observed abundance pattern equally well, the detection of key elements such as C, O, and Al (along with other odd elements) that were not observed in J1010+2358 could be used to distinguish between the two sources. 

\begin{figure}
    \centering
    \includegraphics[width=\columnwidth]{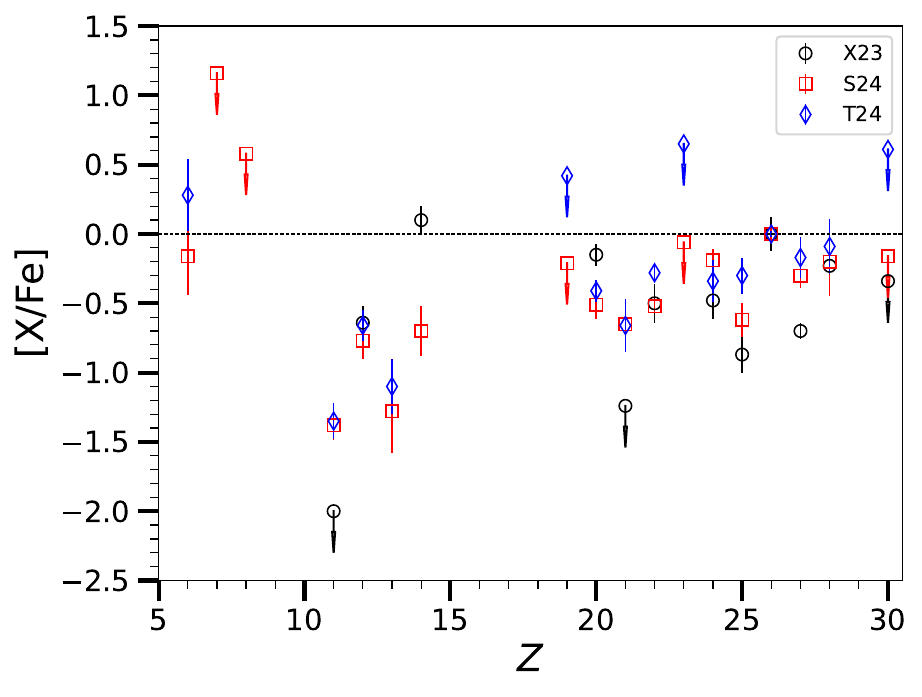}
    \caption{Comparison of the abundance pattern for J1010+2358 from \citetalias{xing2023Natur} (black circles), \citetalias{skuladottir2024ApJ} (red squares), and \citetalias{Thibodeaux2024} (blue diamonds). Note that for \citetalias{skuladottir2024ApJ} we adopt LTE data for Co.}
    \label{fig:data_comparison}
\end{figure}

\citet{Thibodeaux2024} (hereafter \citetalias{Thibodeaux2024}) and \citet{skuladottir2024ApJ} (hereafter \citetalias{skuladottir2024ApJ}) recently independently reanalyzed J1010+2358 with new high-resolution spectra from Keck/HIRES and VLT/UVES, respectively, where they were able to measure the crucial missing elements C and Al. The abundance pattern from these observations is shown in Fig.~\ref{fig:data_comparison} along with the observation from \citetalias{xing2023Natur}. In addition to the new detection of C and Al, Na and Sc were detected by both \citetalias{skuladottir2024ApJ} and \citetalias{Thibodeaux2024} with roughly similar values of $\B{Na}{Fe}\sim -1.4$ and $\B{Sc}{Fe}\sim -0.7$ which are much higher than the upper limit of $\B{Na}{Fe}< -2$ and $\B{Sc}{Fe}<-1.3$ reported by \citetalias{xing2023Natur}. 
\citetalias{skuladottir2024ApJ} also detected a very low Si abundance with highly sub-solar $\B{Si}{Fe}=-0.7$ that is markedly lower than the value of $\B{Si}{Fe}=0.15$ reported by \citetalias{xing2023Natur}. Furthermore, substantial deviation from \citetalias{xing2023Natur} is found for Ca, Mn and Co in the new measurements by \citetalias{skuladottir2024ApJ} and \citetalias{Thibodeaux2024} (see Fig.~\ref{fig:data_comparison}). The new observations by \citetalias{Thibodeaux2024} and \citetalias{skuladottir2024ApJ} are roughly consistent with each other for most elements, especially when the $1\sigma$ error bar is taken into account. As noted by \citetalias{skuladottir2024ApJ}, corrections due to non-local thermodynamic equilibrium (NLTE) effects for Co are rather high and could be uncertain. For this reason,  we adopt the LTE abundance for Co from \citetalias{skuladottir2024ApJ} that is consistent with the NLTE Co abundance estimated by \citetalias{Thibodeaux2024}. 
These new observations essentially rule out the possibility that the abundance pattern in J1010+2358 is from pure PISN ejecta. \citetalias{Thibodeaux2024} concluded that a low-mass $11\,\Msun$ Pop III CCSN can fit the observed abundance pattern well. \citetalias{skuladottir2024ApJ}, on the other hand, find that the abundance pattern can be best explained when the ejecta from a $13\,\Msun$ Pop II star is combined with a faint Pop III CCSN ejecta. \citetalias{skuladottir2024ApJ} also find that when  PISN ejecta combined with a Pop III CCSN or Type 1a SN (SN1a) ejecta combined with a high energy Pop III CCSN, the abundance pattern in J1010+2358 can also be fit, but the quality of fit is poorer and the solution is less favourable. 

Even though the new measurements of J1010+2358 clearly show that the elements are not formed from pure PISN ejecta, the abundance pattern is peculiar compared to normal VMP stars and can be broadly classified as an $\alpha$-poor VMP star that is characterised by sub-solar values $\B{X}{Fe}$ for $\alpha $ elements such as Mg, Si, and Ca. $\alpha$-poor VMP stars have been traditionally associated with a star that formed from gas polluted by SN1a~\citep{ivans2003ApJ, Li2022ApJ}. Such stars could thus help shed light on SN1a contribution to early galactic chemical evolution. However, a recent study by \citet{Jeena_1a_2024arXiv} found that the abundance pattern in some  $\alpha$-poor stars can be explained well by pure CCSN ejecta and does not necessarily need contribution from SN1a.
In order to find the source responsible for the elements observed in J1010+2358, we perform a detailed analysis of the origin of elements observed in J1010+2358  based on the method developed in the recent work by \citet{Jeena_1a_2024arXiv}. We consider a wide range of scenarios for the origin of elements in VMP stars which include ejecta from distinct sources such as PISN, CCSN, and SN1a as well as the combination of ejecta from two such sources.  

\section{Matching the Abundance Pattern from different Theoretical Scenarios }
We follow the procedure for the analysis of the abundance pattern of VMP stars detailed in \citet{Jeena_1a_2024arXiv}. This includes comparing the observed abundance pattern  with the theoretical abundance pattern from six different scenarios;
\begin{enumerate}
    \item Ejecta from a single Pop III PISN.
    \item Ejecta from a single Pop III CCSN that undergoes mixing and fallback.
    \item Combination of ejecta from two Pop III CCSNe, one of which undergoes mixing and fallback. We refer to this as the 2CCSNe scenario.
    \item Combination of ejecta from a single Pop III PISN and a single Pop III CCSN that undergoes mixing and fallback. We refer to this as the CCSN+PISN scenario.
    \item Combination of ejecta from a near-Chandrasekhar mass (near-\ch) SN1a and single Pop III CCSN that undergoes mixing and fallback. We refer to this as the CCSN+near-\ch scenario.
    \item Combination of ejecta from a sub-Chandrasekhar mass (sub-\ch) SN1a and a single Pop III CCSN that undergoes mixing and fallback. We refer to this as the CCSN+sub-\ch scenario.
\end{enumerate}
The PISN yields are adapted from \citet{heger2002} that include models of Pop III stars with He core masses of $65\hbox{--}130\,\Msun$ corresponding to an initial progenitor mass of $\sim 140\hbox{--}260\,\Msun$.
The CCSN yields are adapted from the Pop III models (referred to as the \texttt{z} models) ranging from $10\hbox{--}30\,\Msun$ presented in \citet{jeena_CCSN2024,Jeena_1a_2024arXiv} that are computed using the 1D hydrodynamic code \textsc{kepler} \citep{weaver1978presupernova,rauscher+2003}. For each CCSN model, we adopt two different choices of the mass cut $M_{\rm cut,ini}$ below which all material is assumed to fall back to the proto-neutron star and are named as $Y_{\rm e}$ and $S_4$ models. In the $Y_{\rm e}$ models, $M_{\rm cut,ini}$ is chosen to be at the edge of the Fe core where there is a jump in $Y_{\rm e}$ whereas in $S_4$ models, it is chosen to be at the location where the entropy per baryon exceeds $4k_{\rm B}$. The CCSN models are labelled with the progenitor mass and the choice of $M_{\rm cut,ini}$. For example, a $Y_{\rm e}$ CCSN model from a $25\,\Msun$ Pop III star is labelled as \texttt{z25}-$Y_{\rm e}$.   
For all progenitors, explosion energy $E_{\rm exp}$ of $1.2\times10^{51}\,\erg$ and $1.2\times10^{52}\,\erg$ are adopted. Additionally, $E_{\rm exp }$  of $0.3\times10^{51}\,\erg$ and $0.6\times10^{51}\,\erg$ are also computed for all progenitors of initial mass $< 12\,\Msun$.
The mixing and fallback in CCSN models are parametrized by the parameters  $M_{\rm cut, fin}$ and $f_{\rm cut,}$, where $M_{\rm cut, fin}$ is the mass coordinate above which all material is assumed to be ejected whereas a fraction $f_{\rm cut,}$ is ejected from the material $\Delta M_{\rm cut}=M_{\rm cut,fin}-M_{\rm cut,ini}$ between $M_{\rm cut, ini}$ and $M_{\rm cut, fin}$. The amount of material that falls back to the central remnant following the explosion is $\Delta M_{\rm fb}=(1-f_{\rm cut})\Delta M_{\rm cut}$.

The near-\ch SN1a yields are adopted from the delayed detonation model N100\_Z0.01 by \citet{2013seitenzahl} that corresponds to the explosion of a white dwarf with initial metallicity of $0.01\,{\rm Z_\odot}$ and a central density of $2.9\times10^{9}\,\g\,\cm^{-3}$. The sub-\ch  SN1a yields are adapted from the eleven double detonation models presented in \citet{2021gronow} with an initial metallicity of $0.001\,{\rm Z_\odot}$ of white dwarf with CO core masses of $0.8$--$1.1\,\Msun$ and He shell masses of $0.02$--$0.1\,\Msun$. The models are labelled using the values of CO core and He shell mass. For example, M10\_05 refers to a CO core mass of $1.0\,\Msun$ and He shell mass of $0.05\,\Msun$.

In scenarios 3--6 listed above which involve combining the ejecta from CCSN with mixing and fallback with another source S2, the mixing is parameterized by a single parameter $\alpha$ given by
\begin{equation}
    \alpha =\frac{ M{\rm_{dil,CCSN}}}{M{\rm_{dil,CCSN}}+M{\rm_{dil,S2}}},
    \label{eq:alpha}
\end{equation}
where $M{\rm_{dil,CCSN}}$ and $M{\rm_{dil,S2}}$ are the effective dilution masses from CCSN and source S2, respectively. We impose a minimum value of dilution mass of $10^4\,\Msun$ for SN1a as well as all CCSN models with explosion energy $\leq1.2\times10^{51}\,\erg$. For high-energy CCSN and PISN, we impose a minimum dilution of $10^5\,\Msun$. The best-fit model from each scenario is found by minimizing the $\chi^2$ as described in \citet{heger2010nucleosynthesis,jeenaCEMP2023}. For any given fit from scenarios with two sources S1 and S2, the relative contribution from each source to an element ${\rm X}_i$, can be quantified by computing the fraction $\eta ({\rm X}_i)$ of the total elemental yield $Y_{{\rm X}_i}$ where $\eta_{\rm S1}({\rm X}_i)+\eta_{\rm S2}({\rm X}_i)=1$. We note that except for the scenario with pure PISN ejecta i.e., scenarios that involve CCSN, we treat Sc as an upper limit. This is because Sc is dominantly produced by neutrino-processed proton-rich ejecta that is not modelled in our 1D calculations \citep{siverding2020ApJ,wang2024ApJ}. 

We note that we do not explicitly consider the scenario of pure SN 1a ejecta. This is because the chance of pristine gas polluted purely by SN 1a is exceedingly unlikely due to the inherent delay times associated with SN 1a. CCSN would invariably contribute at some level before SN 1a can contribute to the enrichment of the gas in the early Galaxy. Nevertheless, because we consider varying proportions of SN 1a contribution in evaluating the best-fit CCSN+SN 1a model, almost pure SN 1a is also considered as one of the possibilities corresponding to $\alpha\approx 1$. For the same reason, the scenario of combining ejecta from a single PISN and SN 1a is not considered as it is extremely unlikely since CCSN would invariably enrich the gas in the early Galaxy before SN 1a.

\section{Best-fit Models for J1010+2358}
\subsection{Analysis of \citetalias{xing2023Natur} data}
\begin{figure}
    \centering
    \includegraphics[width=\columnwidth]{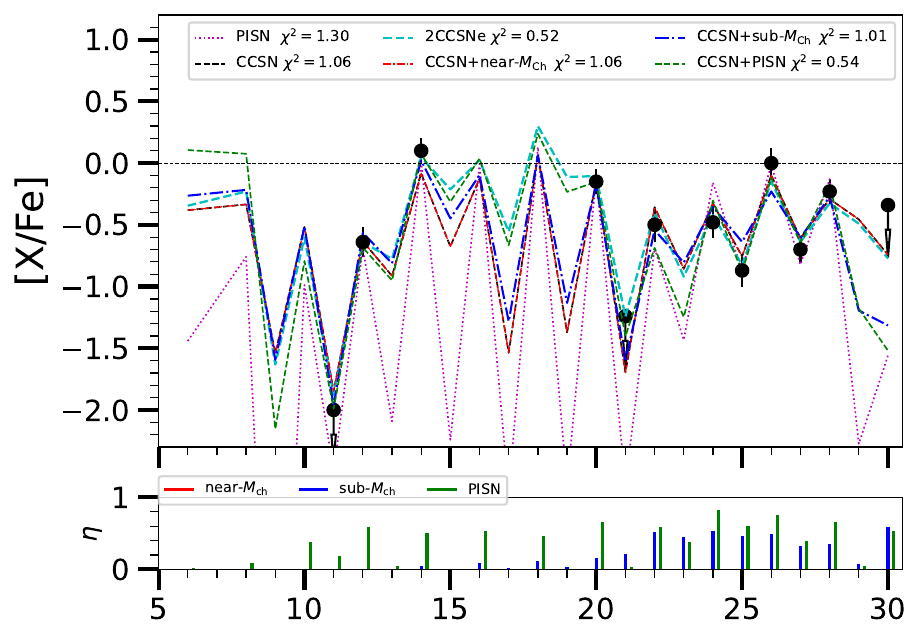}
    \caption{Top: The elemental abundance pattern of J1010+2358 from \citetalias{xing2023Natur}, compared with the best-fit models from various scenarios: PISN (magenta dotted line), CCSN (black dashed line), 2CCSNe (cyan dashed line),  CCSN+near-\ch (red dash-dotted line), CCSN+sub-\ch (blue dash-dotted line), and CCSN+PISN (green dashed line). Bottom: The red, blue, and green vertical solid lines show the fraction of elements $\eta_{\rm 1a}$ produced by near-\ch SN1a model, sub-\ch SN1a model, and PISN model, respectively.}
    \label{fig:best-fit_xing}
\end{figure}

We first analyse the abundance pattern from the earlier detection by \citetalias{xing2023Natur} as a starting point followed by the two recent observations by \citetalias{Thibodeaux2024} and \citetalias{skuladottir2024ApJ}. Figure~\ref{fig:best-fit_xing} shows the best-fit models for all the scenarios for \citetalias{xing2023Natur} and the best-fit parameters are listed in Table~\ref{tab:best-fit_parm}. As can be seen from the figure, the best-fit models from \emph{all six} scenarios provide a good fit to the observed abundance pattern with the lowest $\chi^2=0.52$ from the 2CCSNe scenario and the highest $\chi^2=1.30$ from the single PISN scenario.  However, as can be seen from the figure, the best-fit 2CCSNe model provides a near-perfect fit where it can match the abundances of all elements within the $1\sigma$ uncertainty in contrast to the best-fit single PISN model that fails to match Ti and Cr. 
The best-fit model from the CCSN+PISN scenario also provides an excellent fit with $\chi^2=0.54$ that is almost comparable to the best-fit model from the 2CCSNe scenario although it fails to match Ti abundance within the $1\sigma$ uncertainty.  Interestingly, the CCSN model in the best-fit model from the CCSN+PISN scenario is from an intermediate-mass CCSN model of $19.4\,\Msun$ with a relatively large fallback mass of $2.2\,\Msun$. This is distinct from the best-fit models from all other scenarios involving CCSN that have low mass CCSN models of $\sim 13\,\Msun$ with negligible fallback. The PISN in the best-fit CCSN+PISN scenario contributes considerably ($\gtrsim 50\%$) to most even $Z$ elements from Ne to Zn as can be seen from values of $\eta$ in the bottom panel of Fig.~\ref{fig:best-fit_xing}. The best-fit model from the CCSN+near-\ch scenario is identical to the single CCSN scenario with the same $12.8\,\Msun$ CCSN model. The contribution from SN1a is negligible with $\eta_{\rm 1a}\lesssim 10^{-5}$ for all elements. In contrast, the best-fit model from CCSN+sub-\ch scenario has a distinct CCSN model with a substantial contribution from SN1a with $\eta_{\rm 1a}\sim 20\hbox{--}50\%$ for most elements from Ca to Zn.

\subsection{Analysis of \citetalias{skuladottir2024ApJ} data}
\begin{figure}
    \centering
    \includegraphics[width=\columnwidth]{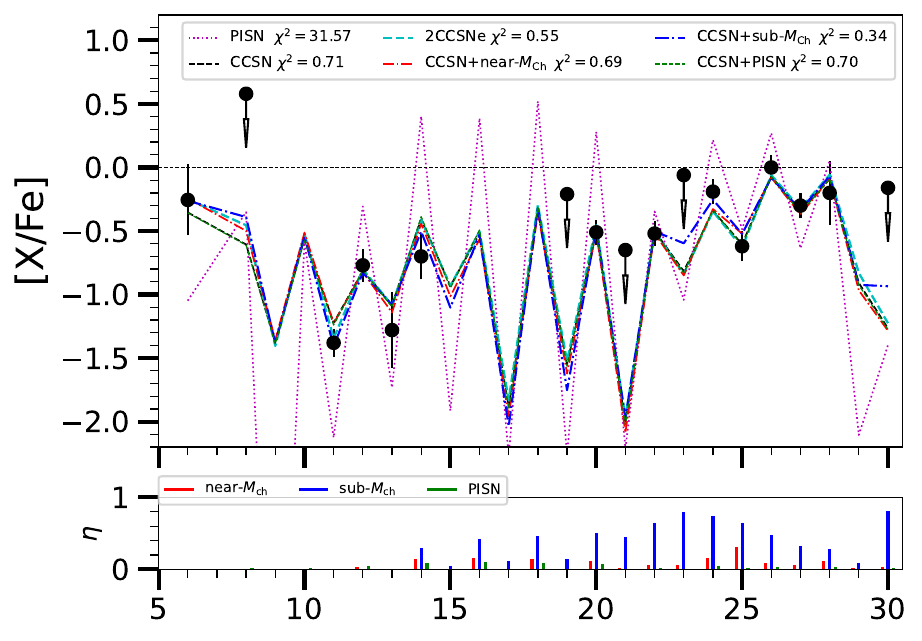} \\
    \caption{Same as Fig.~\ref{fig:best-fit_xing}, but the data is from \citetalias{skuladottir2024ApJ}.}
    \label{fig:best_fit_asask}
\end{figure}

We now consider the abundance pattern measured by \citetalias{skuladottir2024ApJ}. Figure~\ref{fig:best_fit_asask} shows the best-fit models from all the scenarios compared to the observed abundance pattern. Unlike the \citetalias{xing2023Natur} data, pure PISN ejecta provides an extremely poor fit as is evident from the very high $\chi^2=31.6$. All the other five scenarios that involve CCSN provide a very good fit to the observed abundance pattern and can match almost all elements within the $1\sigma$ observed uncertainty. The best-fit CCSN+sub-\ch model has the lowest $\chi^2=0.34$ and is the only model that can match the abundance of all the observed elements within the $1\sigma$ uncertainty. 
In this case, the ejecta from M10\_10 sub-\ch model contributes to all elements above Si and dominantly to Ti--Mn and Zn where the rest comes from an intermediate-mass CCSN of $15.2\,\Msun$ with a somewhat large fallback of $\sim 1\,\Msun$.   
Although the lowest $\chi^2$ is from the best-fit CCSN+sub-\ch model, it is clear that low mass CCSN with low fallback from the best-fit 2CCSNe and single CCSN scenarios can provide an excellent fit with comparable $\chi^2$ of $0.55$ and $0.71$, respectively. The best-fit CCSN+near-\ch model can also provide a very good fit with $\chi^2=0.69$ but has a negligible contribution from SN1a with $\eta_{\rm 1a}\lesssim 0.15$ for all elements except Mn that has $\eta_{\rm 1a}=0.3$. The CCSN model in this case is also identical to one of the models from the best-fit 2CCSNe scenario. Finally, the best-fit CCSN+PISN model is essentially the same as the best-fit single CCSN model with almost identical $\chi^2$ and the same CCSN progenitor. The contribution from PISN is negligible with $\eta_{\rm PISN}\lesssim 0.1$ for all detected elements. This reconfirms the fact that PISN is entirely unnecessary to fit the abundance pattern and the pattern has no particular PISN features. 

\begin{figure}
    \centering
    \includegraphics[width=\columnwidth]{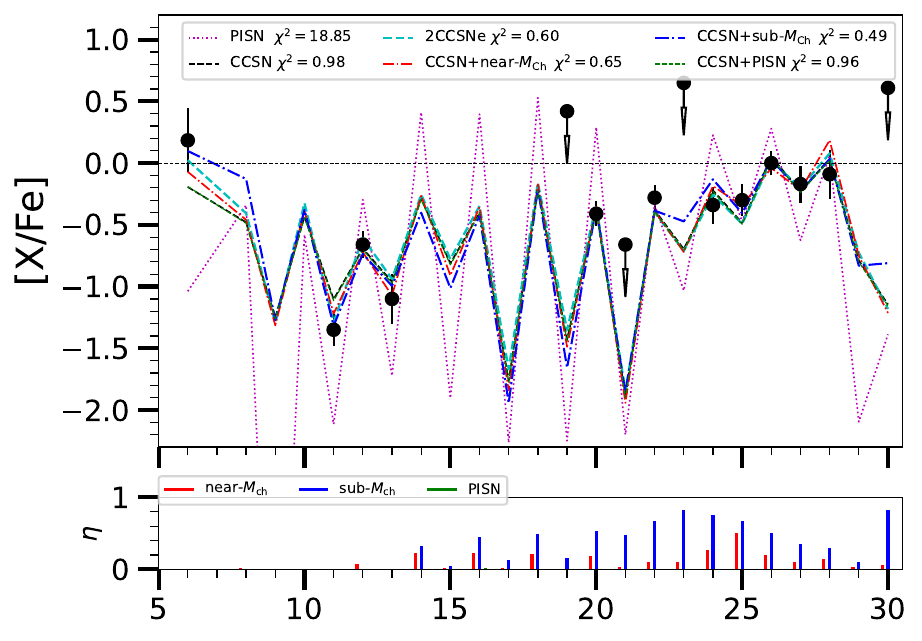}
    \caption{Same as Fig.\ref{fig:best-fit_xing}, but the data is from \citetalias{Thibodeaux2024}.}
    \label{fig:best_fit_pierre}
\end{figure}

\begin{table*}
    \centering
    \caption{Best-fit models and parameters along with the dilution mass. }
    \begin{tabular}{ccccccccc}
    \hline
    Star&Scenario&Model name&$E_{\rm exp}$ &$\chi^2$&$\alpha$&$\Delta M_{
    \rm cut}$&$\Delta M_{\rm fb}$&$M{\rm_{dil}}$\\
    && & ($\times 10^{51}\,\erg$)& &&(\Msun) &(\Msun)& ($\times 10^4\,\Msun$)\\
    \hline
    \multirow{9}{*}{J1010+2358}&  PISN& $130\,\Msun$ He core&87.3&1.30&--&--&--&$1.1\times 10^5$\\ 
       \multirow{9}{*}{\citetalias{xing2023Natur}} &CCSN& \texttt{z12.8}-$S_{4}$ &1.2&1.06&--& 0.10&0.02 &2.24\\ 
        &\multirow{2}{*}{CCSN+PISN}& \texttt{z19.4}-$Y_{\rm e}$+&1.2&\multirow{2}{*}{0.54}&\multirow{2}{*}{$9 \times 10^{-4}$}&2.31&2.17&1.24\\
                                &  &$130\,\Msun$ He core&87.3&&&--&--&$1.4 \times 10^3$\\
         &\multirow{2}{*}{2CCSNe} & \texttt{z12.8}-$S_{4} +$&1.2&\multirow{2}{*}{ 0.52}&\multirow{2}{*}{0.68}& 0.00&0.00&2.91 \\
         && \texttt{z13}-$S_{4} $&1.2 & &&0.01 &0.01 &6.18\\ 
          &\multirow{2}{*}{CCSN+near-\ch}&\texttt{z12.8}-$S_{4} +$&1.2 &\multirow{2}{*}{1.06}&\multirow{2}{*}{$ 10^{-7}$}& 0.10&0.02 &2.24\\ 
         &&  N100\_Z0.01&-- & &&-- &--&$ 10^7$ \\ 
         &\multirow{2}{*}{CCSN+sub-\ch}&\texttt{z13.2}-$Y_{\rm e} +$& 1.2&\multirow{2}{*}{1.01}&\multirow{2}{*}{0.03}& 0.30&0.26 &1.92\\
         &&   M11\_05& --&& &-- &--&62.1 \\
         \hline
   \multirow{9}{*}{J1010+2358}&  PISN& $125\,\Msun$ He core&78.8&31.57&--&--&--&$4.5\times 10^4$\\
    \multirow{9}{*}{\citetalias{skuladottir2024ApJ}}&CCSN& \texttt{z11.8}-$Y_{\rm e}$ &1.2&0.71&--& 0.50&0.36 &1.00\\
     &\multirow{2}{*}{CCSN+PISN}& \texttt{z11.8}-$Y_{\rm e}$+&1.2&\multirow{2}{*}{0.70}&\multirow{2}{*}{$4\times 10^{-5}$}&0.50&0.37&1.00\\
         &&$130\,\Msun$ He core&87.3&&&--&--&$2.5 \times 10^4$\\
    &\multirow{2}{*}{2CCSNe} & \texttt{z13.5}-$Y_{\rm e} +$ &1.2&\multirow{2}{*}{0.55}&\multirow{2}{*}{0.08}&0.00&0.00&13.7\\
    &&\texttt{z11.6}-$Y_{\rm e}$&1.2& & & 0.39& 0.30& 1.19\\
    &\multirow{2}{*}{CCSN+near-\ch}&\texttt{z11.6}-$Y_{\rm e} +$&1.2&\multirow{2}{*}{0.69}&\multirow{2}{*}{0.005}& 0.39&0.28 & 1.01\\
    &&  N100\_Z0.01& --& && --&-- &202\\ 
    &\multirow{2}{*}{CCSN+sub-\ch}&\texttt{z15.2}-$Y_{\rm e} +$& 1.2&\multirow{2}{*}{0.34}&\multirow{2}{*}{0.05}& 1.17&0.97 &2.23\\
    &&   M10\_10& --&& &-- &--&42.3 \\
         \hline
   \multirow{9}{*}{J1010+2358}&  PISN& $125\,\Msun$ He core&78.8&18.85&--&--&--&$6.7\times 10^4$\\
    \multirow{9}{*}{\citetalias{Thibodeaux2024}}&CCSN& \texttt{z11.8}-$Y_{\rm e}$ &1.2&0.98&--& 0.50&0.37 &1.04\\
    &\multirow{2}{*}{CCSN+PISN}& \texttt{z11.8}-$Y_{\rm e}$+&1.2&\multirow{2}{*}{0.96}&\multirow{2}{*}{$10^{-6}$}&0.50&0.38&1.00\\
         &&$120\,\Msun$ He core&71.0&&&--&--&$2.5 \times 10^5$\\
    &\multirow{2}{*}{2CCSNe} & \texttt{z11.9}-$Y_{\rm e} +$ &1.2&\multirow{2}{*}{0.60}&\multirow{2}{*}{0.11}&0.00&0.00&8.16\\
    &&\texttt{z13.4}-$Y_{\rm e}$&1.2& & & 0.78&0.72 & 1.01\\
    &\multirow{2}{*}{CCSN+near-\ch}&\texttt{z12.1}-$Y_{\rm e} +$&1.2&\multirow{2}{*}{0.65}&\multirow{2}{*}{0.008}& 0.50&0.41 & 1.02\\
    &&  N100\_Z0.01&-- & &&-- &-- &127\\ 
    &\multirow{2}{*}{CCSN+sub-\ch}&\texttt{z15.2}-$Y_{\rm e} +$& 1.2&\multirow{2}{*}{0.49}&\multirow{2}{*}{0.03}& 1.17&1.06 &1.45\\
    &&   M10\_10& --&& &-- &--&46.8 \\
         \hline
    \end{tabular}
    
    \label{tab:best-fit_parm}
\end{table*}

\subsection{Analysis of \citetalias{Thibodeaux2024} data}
The situation is very similar when the observed abundance pattern from \citetalias{Thibodeaux2024} is considered. The best-fit models from all scenarios are shown in Fig.~\ref{fig:best_fit_pierre} with the corresponding details listed in Table~\ref{tab:best-fit_parm}. Pure PISN fails to fit the abundance pattern with a very high $\chi^2=18.9$ whereas all other scenarios involving CCSN provide an excellent fit. The best-fit CCSN+sub-\ch model has the lowest $\chi^2=0.49$ with the same combination of CCSN and sub-\ch SN1a model as the best-fit model that matches \citetalias{skuladottir2024ApJ}. Low mass CCSN with negligible fallback can provide an excellent fit from both the single CCSN and 2CCSNe scenario with $\chi^2$ of $0.98$ and $0.60$, respectively. The best-fit CCSN+near-\ch  model provides a very good fit with $\chi^2=0.65$ but SN1a makes a substantial contribution to only Mn with the rest coming from a low-mass CCSN of $12.1\,\Msun$. The quality of fit from the best-fit CCSN+PISN model is identical to the best-fit single CCSN model with near zero contribution from PISN with $\eta_{\rm PISN}\lesssim 0.01$ for all elements which is even lower than in the case of \citetalias{skuladottir2024ApJ}. This again clearly indicates that the abundance pattern has no PISN features.

\section{Comparison with previous analysis}
\subsection{Comparison with \citet{jeena_CCSN2024}}
Our analysis of  J1010+2358 from the \citetalias{xing2023Natur} data is consistent with the results found by \citet{jeena_CCSN2024} that although pure PISN can fit the observed abundance pattern, single CCSN models can provide slightly better fits. Because the analysis performed in this study is much more exhaustive compared to \citet{jeena_CCSN2024}, we are able to explore a larger set of scenarios. We find that although all scenarios provide a good fit, the best-fit 2CCSNe model provides a near-perfect fit to all elements which is substantially better than the best-fit model from the pure PISN scenario. This is further evidence that the measured pattern by \citetalias{xing2023Natur} can be fit better by CCSN. We note that the CCSN+PISN scenario also provides a fit comparable to the 2CCSNe scenario. However, because of the mixing of CCSN and PISN ejecta, the key PISN feature of strong odd-even effect is absent in the CCSN+PISN models making them almost indistinguishable from the CCSN or 2CCSNe fits.

\subsection{Comparison with \citetalias{skuladottir2024ApJ} and \citetalias{Thibodeaux2024}}
With regard to the analysis by \citetalias{skuladottir2024ApJ} and \citetalias{Thibodeaux2024} based on their new measurements, our results are consistent with their main conclusion that pure PISN ejecta is incompatible with the new abundance pattern. Both \citetalias{skuladottir2024ApJ} and \citetalias{Thibodeaux2024} find that the abundance pattern can be explained well purely by CCSN ejecta. Our analysis also finds that both single CCSN and 2CCSNe scenarios can explain the new abundance pattern very well. \citetalias{skuladottir2024ApJ} found that the abundance pattern can also be fit by combining PISN and CCSN ejecta but the quality of fit compared to the best-fit becomes progressively worse with increasing PISN contribution. This is also borne out in our analysis where the maximum PISN contribution to any element in the best-fit CCSN+PISN model is $\lesssim 10\%$ and $\lesssim 1\%$ for the  \citetalias{skuladottir2024ApJ} and \citetalias{Thibodeaux2024} data, respectively.

The results from our analysis have some important differences particularly when compared to the results found in \citetalias{skuladottir2024ApJ}. Unlike in \citetalias{skuladottir2024ApJ} where a good fit from single Pop III CCSN models was not found, we find an excellent fit to their observed pattern from a Pop III $11.8\,\Msun$ model. In contrast, most of the best-fit solutions found by \citetalias{skuladottir2024ApJ} require contribution from a $13\,\Msun$ Pop II CCSN model from \citet{woosley1995ApJS}.  
The difference in the result is likely because \citetalias{skuladottir2024ApJ} included Sc in their analysis along with older CCSN models. As noted in \citet{jeena_CCSN2024}, the Sc in Pop III and Pop II in all of the latest CCSN models from \textsc{kepler} is very low with $\B{Sc}{Fe}\lesssim -1$ with the exception of models that experience merger of the O burning shell with the O-Ne-Mg shell during the pre-SN evolution. However, such models that undergo shell mergers have super-solar values of $\B{X}{Fe}$ for Si and Ca along with enhanced Al and are thus incompatible with the observations. This is in contrast to the $13\,\Msun$ Pop II CCSN model from \citet{woosley1995ApJS} where a higher $\B{Sc}{Fe}\sim -0.6$ was found. The situation is similar for \citetalias{Thibodeaux2024} who also included Sc and found the $11\,\Msun$ Pop III CCSN model to be the best fit which happens to also provide $\B{Sc}{Fe}\sim -0.6$ in contrast to our CCSN models. 
As noted earlier, Sc is copiously produced in the neutrino-processed ejecta during the CCSN explosion and is not included in 1D models. For this reason, as in our analysis, Sc should always be treated as an upper limit when analysing VMP stars when using yields from 1D models. 

In addition to the mixing of PISN and CCSN ejecta, \citetalias{skuladottir2024ApJ} also considered the mixing of CCSN and SN 1a ejecta and found that the best fit from such a combination is considerably worse. This is in contrast to our results where the overall best fit is from the CCSN+sub-\ch scenario while the CCSN+near-\ch scenario also provides an excellent fit. With regard to CCSN+sub-\ch models, other than the difference in our treatment of Sc mentioned above, the simple reason is that \citetalias{skuladottir2024ApJ} only considered near-\ch SN 1a models and did not consider sub-\ch SN 1a models. 
With regard to our CCSN-near-\ch models, the reason for the difference from \citetalias{skuladottir2024ApJ} (other than treatment of Sc) cannot be attributed to SN 1a models as the near-\ch SN 1a model N100\_Z0.01 considered in this study from \citet{2013seitenzahl} is very similar to the model by \citet{iwamoto1999} adopted in \citetalias{skuladottir2024ApJ}.  The difference is likely because our large set of CCSN models, along with mixing and fallback, cover much larger combinations of CCSN ejecta than what is considered by \citetalias{skuladottir2024ApJ}. The much larger set of CCSN ejecta models when combined with near-\ch SN 1a ejecta are likely to find better fits. We note that our best-fit CCSN+near-\ch model has a negligible contribution from SN 1a compared to $40\%$ contribution found in \citetalias{skuladottir2024ApJ}.

\section{Summary and Conclusion}
We find that for the updated abundance pattern from \citetalias{skuladottir2024ApJ} and \citetalias{Thibodeaux2024}, the best-fit model from CCSN+sub-\ch scenario provides the overall best-fit which is closely followed by 2CCSNe, CCSN+near-\ch, and single CCSNe. We find that although the best-fit CCSN+PISN model can also provide a very good fit, it has negligible PISN contribution which further indicates the lack of PISN features in the observed abundance pattern.  

In conclusion, the updated measurements of the abundance pattern in the chemically peculiar VMP star J1010+2358 by \citetalias{skuladottir2024ApJ} and \citetalias{Thibodeaux2024} clearly rule out the star as being born from a gas polluted purely by a PISN. Instead, the pattern can be explained by purely CCSN ejecta or a combination of CCSN and SN1a ejecta. Thus, a VMP star born from purely PISN ejecta remains yet to be identified and will likely require measurement of detailed abundance patterns in a much larger number of VMP stars.

\section*{Data Availability}
Data is available upon reasonable request.

\bibliography{main}{}
\bibliographystyle{aasjournal}
\end{document}